\documentclass{article} 
\usepackage{iclr2027_conference,times}

\usepackage{amsmath,amsfonts,bm}

\def\eqref#1{equation~\ref{#1}}

\def\1{\bm{1}}

\DeclareMathAlphabet{\mathsfit}{\encodingdefault}{\sfdefault}{m}{sl}
\SetMathAlphabet{\mathsfit}{bold}{\encodingdefault}{\sfdefault}{bx}{n}

\usepackage{hyperref}
\usepackage{url}

\usepackage{graphicx} 
\usepackage{tabularx}
\usepackage{booktabs}
\usepackage{placeins}
\usepackage{amsmath}
\usepackage{dsfont}
\usepackage{multirow}
\usepackage{array}
\usepackage{makecell}
\usepackage{adjustbox}
\usepackage{enumitem}

\usepackage{amssymb}
\usepackage{wrapfig}

\title{RISR: Residual-Informed Scientific Equation Discovery with Large Language Models}

\author{%
\noindent
\textbf{Haobo Li}\textsuperscript{1},
\textbf{Wenshuo Zhang}\textsuperscript{2},
\textbf{Wenxiao Zhao}\textsuperscript{3},
\textbf{Eunseo Jung}\textsuperscript{2},
\textbf{Rui Sheng}\textsuperscript{2},
\textbf{Yushi Sun}\textsuperscript{2},\\
\textbf{Peiqin Zhuang}\textsuperscript{1},
\textbf{Hao Chen}\textsuperscript{1},
\textbf{Fenghua Ling}\textsuperscript{1}\\
\textsuperscript{1}Shanghai AI Laboratory
\quad
\textsuperscript{2}The Hong Kong University of Science and Technology
\\
\textsuperscript{3}University of California, Los Angeles\\
}

\iclrfinalcopy 
\begin{document}

\newcommand{\name}{RISR}

\maketitle

\begin{abstract}
Symbolic regression combines structural search with numerical fitting, but aggregate fit scores do not describe how the remaining error varies across inputs. We introduce RISR, a residual-informed method that uses these error patterns to guide formula discovery and learn which corrections are worth fitting. A residual encoder compresses aligned inputs, targets, current predictions, and residuals into continuous tokens that condition a language model to propose formulas. 
For subsequent refinement, a dual-view relational encoder uses additive and regularized multiplicative residuals to predict the post-fit utility of candidate corrections. 
We evaluate RISR on scientific tasks from the LLM-SRBench. RISR achieves 63.57\% and 38.50\% ID accuracy at the 1\% and 0.1\% pointwise relative-error tolerances, respectively. The corresponding OOD accuracies are 56.07\% and 38.24\%. \name{} outperforms the reported baselines using the same backbone. The results show that our residual-informed approach can improve numerical equation recovery.

\end{abstract}

\section{Introduction}
\label{sec:introduction}
Symbolic regression (SR) discovers explicit mathematical relationships from observations, providing interpretable models for scientific analysis~\citep{schmidt2009distilling,udrescu2020aifeynman,cranmer2023pysr}. These expressions make dependencies between variables explicit and allow proposed relationships to be examined and compared. Discovering an equation requires both discrete search over expression structures and continuous estimation of numerical coefficients. The two are closely coupled: the usefulness of a structural term depends on its fitted coefficients and its combination with the rest of the equation. A discovery procedure must therefore explore candidate structures while accounting for their numerical behavior on the observations.

Existing methods address this coupled problem through different representations and search strategies.
Sparse identification selects a small set of terms from a candidate function library~\citep{brunton2016discovering}. Neural expression search learns a sampling policy through risk-seeking policy gradients~\citep{brenden2021deep}, while hybrid frameworks integrate neural search, genetic programming, and other complementary strategies~\citep{landajuela2022unified}. Another line of work trains expression generators on synthetic equation--observation pairs, learning to map numerical observations to symbolic expressions~\citep{biggio2021neural,kamienny2022endtoend}. Together, these approaches provide mechanisms for proposing structures and organizing exploration of the expression space.

Recent work increasingly uses numerical evaluation to guide large language models (LLMs) during search. TPSR incorporates fitting accuracy and expression complexity into Transformer decoding through Monte Carlo tree search~\citep{shojaee2023transformer}. For LLM-based program generation, FunSearch combines a pretrained language model with an evaluator in an evolutionary search procedure~\citep{romera2024mathematical}. LLM-based equation-discovery systems use scientific knowledge and executable programs to construct and revise candidate relationships~\citep{grayeli2024lasr,shojaee2025llmsr,xia2026scientist}. LLM-SR scores fitted candidates by their mean squared error and uses a score-organized experience buffer to construct subsequent prompts~\citep{shojaee2025llmsr}. SR-Scientist further allows the model to analyze data and residual errors through code execution~\citep{xia2026scientist}.

The form of numerical feedback determines what guidance is available for the next proposal. Aggregate fit scores rank candidate equations, but do not preserve the location or variable dependence of their errors.
Equations with comparable overall losses can therefore require different corrections.
Pointwise residuals expose how the remaining error varies across inputs, including nonlinear patterns, variable interactions, and localized discrepancies. Turning this information into useful search guidance requires accounting for the current equation and the effects of numerical fitting.
A correction suggested by a residual pattern may produce different improvements depending on its parameterization and how it is combined with the current equation.
We study how to encode these numerical states for formula generation and how to learn which corrections are worth fitting.

We introduce \textbf{R}esidual-\textbf{I}nformed \textbf{S}ymbolic \textbf{R}egression (\name{}), a method for scientific equation discovery.
Its search encoder compresses aligned inputs, targets, predictions, and residuals into continuous tokens that condition LLM proposals~\citep{li2021prefix}. 
These proposals are complemented by
a learned structural-edit generator. Candidate formulas are fitted and evaluated to obtain an initial equation.
For subsequent refinement, a dual-view encoder captures variable-specific patterns and interactions from additive and regularized multiplicative residuals.
It models relationships between variables through attention, following the principle of attention-based set encoding~\citep{lee2019settransformer}.
A scoring head uses these representations to predict the utility of candidate corrections after coefficient fitting. Corrections are fitted numerically and accepted when the training loss improves, after which the residuals and action ranking are refreshed.

The encoders are trained separately using synthetic data constructed from randomly transformed and composed elementary functions, independently of benchmark equations and observations.
The search encoder and its projection are trained from scratch on residual-to-formula examples while the LLM remains frozen.
The refinement encoder and scoring head learn from paired current equations with shared observations and candidate actions. Corrections are fitted on one subset of observations, and their outcomes on a disjoint subset provide utility supervision.

We evaluate \name{} on scientific problems in the released LLM-SRBench data, covering materials science, chemistry, biology, and physics~\citep{shojaee2025bench}.
With Qwen3-Coder-30B-A3B-Instruct as the LLM backbone, \name{} achieves 63.57\% and 38.50\% accuracy at the 1\% and 0.1\% relative-error tolerances.
Its performance exceeds that of existing non-LLM or LLM-based methods with the same LLM as backbone and remains competitive with substantially larger backbones.
Our contributions are summarized as follows:
\begin{itemize}[itemsep=0pt, topsep=0pt, parsep=0pt]
    \item We introduce \name{}, a residual-informed method that guides both formula search and subsequent refinement. A search encoder conditions LLM proposals on numerical error patterns, while a dual-view refinement encoder supports correction selection.

    \item We construct complementary synthetic datasets for training \name{}. Random transformations and compositions generate residual-to-formula examples, while paired current equations with shared observations and candidate actions provide fitted-utility supervision on observations held out from coefficient fitting.

    \item We evaluate \name{} on scientific problems. Our method improves overall accuracy over the reported 30B baselines and achieves the highest overall ID accuracy at the 1\% tolerance among the baselines, including those using larger backbones.
\end{itemize}

\vspace{-12pt}
\section{Related work}
\label{sec:related}

\paragraph{Symbolic equation discovery}

Symbolic regression combines expression search with numerical evaluation to recover mathematical relationships from data~\citep{schmidt2009distilling}. To make this search more efficient, physics-inspired methods such as AI Feynman exploit properties including symmetry and separability to simplify the search~\citep{udrescu2020aifeynman}. PySR uses a multi-population evolutionary procedure that alternates expression evolution, simplification, and coefficient optimization~\citep{cranmer2023pysr}. Moving beyond explicit search procedures, neural approaches learn from generated equations and observations: NeSymReS pretrains a Transformer to predict symbolic expressions, while end-to-end symbolic regression predicts expressions and constants jointly before numerical refinement~\citep{biggio2021neural,kamienny2022endtoend}.
TPSR incorporates fitting accuracy and expression complexity into Transformer decoding through Monte Carlo tree search~\citep{shojaee2023transformer}.

The score of a fitted equation does not determine which structural correction will improve it.
The outcome also depends on the current residual pattern, the correction's parameterization, and its composition with the equation.
\name{} addresses this decision by learning correction utility from fitted outcomes.
Its dual-view encoder represents additive and regularized multiplicative correction signals, while a shared scorer uses these representations to rank corrections before fitting.

\paragraph{LLM-based Symbolic Regression}
LLM-based methods combine program generation with executable evaluation.
FunSearch places a pretrained LLM and an evaluator within an evolutionary program-search loop~\citep{romera2024mathematical}.
LLM-SR combines language-model proposals, parameter fitting, and evolutionary search for SR~\citep{shojaee2025llmsr}. 
LaSR learns a textual concept library from high-performing hypotheses and uses it to guide subsequent search~\citep{grayeli2024lasr}. Alternatively, SR-Scientist leverages agents for data analysis, equation evaluation, and iterative optimization, including residual inspection through code
execution~\citep{xia2026scientist}. 

Beyond making residuals available as agents' input, a central question is how to represent the current numerical state for subsequent formula proposals.
\name{} learns this representation from synthetic residual-to-formula examples.
It jointly processes aligned inputs, targets, current predictions, and residuals, compressing them into fixed-size continuous tokens that condition a frozen LLM alongside the textual search context.

\section{RISR}
\label{sec:method}

\begin{figure}[!htbp]
    \centering
    \includegraphics[width=\textwidth]{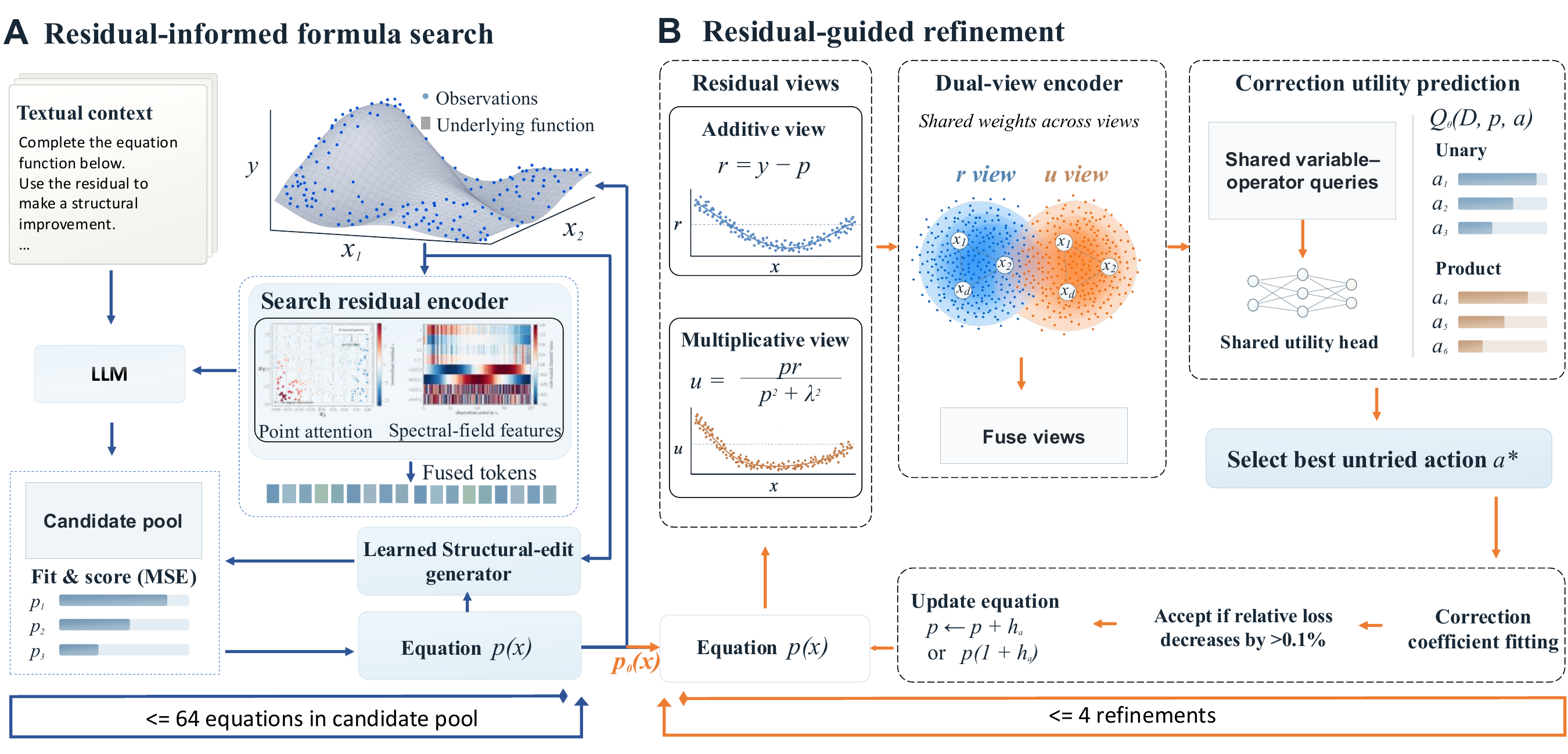}
    \caption{Overview of RISR. (A) Residual-informed formula search conditions a frozen LLM on textual context and numerical tokens from the search residual encoder. LLM proposals and learned structural edits are fitted and evaluated to select an initial equation $p_0$. (B) Residual-guided refinement encodes additive and regularized multiplicative residual views and predicts the fitted utility of unary and product corrections.}
    \label{fig:framework}
    \vspace{-12pt}
\end{figure}

Given a set of training observations $\mathcal D=\{(x_i,y_i)\}_{i=1}^{n}$, where $x_i\in\mathbb R^d$ contains the input variables of observation $i$ and $y_i\in\mathbb R$ is its observed response, RISR seeks an explicit equation that predicts the response from the inputs.
As shown in Figure~\ref{fig:framework}, \name{} proceeds
in two stages. 
In stage A, residual-informed formula search uses a search residual encoder to supply pointwise numerical feedback to a frozen LLM through continuous tokens. LLM proposals and learned structural edits are numerically fitted and evaluated to obtain an initial equation $p_0$.
In stage B, residual-guided refinement then uses a dual-view refinement encoder and utility head to rank correction actions for the current equation. Selected corrections are fitted and accepted when they improve the training loss.

\subsection{Residual-Informed Formula Search}
\label{sec:upstream-encoder}

Inspired by recent work on LLM-guided program search and symbolic regression~\citep{romera2024mathematical,grayeli2024lasr,shojaee2025llmsr}, we iteratively generate and evaluate equation programs. 
The LLM receives two complementary inputs: textual context describing previously evaluated equations and numerical tokens representing a fitted parent's residual state.

\textbf{Textual context.} Each prompt contains equation programs sampled from the candidate pool. The highest-scoring equation included in the prompt is the parent equation used for numerical conditioning.
The textual context provides the equation-completion instruction, variable descriptions, and sampled programs.

\textbf{Numerical residual.} The search residual encoder processes aligned tuples $(x_i,y_i,p(x_i),y_i-p(x_i))$.
A point-attention path aggregates these observations, while a spectral-field path represents coordinate--residual interactions
using frequency features and residual-region pooling.
Learned gates fuse the two paths, and a projection maps their
outputs to 16 continuous tokens in the LLM embedding space.
These tokens accompany the textual context, conditioning executable
formula proposals on the parent's input-dependent error pattern.
A learnable structural-edit generator also supplies candidates. Given the observations and current formula, the generator combines its own residual encoder with an encoding of the formula's abstract syntax tree (AST) to predict an addition, replacement, deletion, or stop action.
The resulting equation $p_0$, with the lowest mean squared error (MSE), initializes downstream refinement.
Appendix~\ref{app:upstream-encoder} gives more details of the encoder architecture and proposal procedures.

\subsection{Residual-Guided Refinement}
Stage B constructs residual views, scores candidate actions, and fits the selected correction.

\subsubsection{Dual-View Residual Encoding}
For a current equation $p$, let $p_i=p(x_i)$ be its prediction and $r_i=y_i-p_i$ its residual. 
We consider two ways to correct its predictions.
Let $h(x)$ denote a candidate correction.
1) An additive update, $p+h$, adds an output increment, so matching the observations requires $h(x_i)\approx r_i$.
2) A multiplicative update, $p(1+h)$, applies a relative adjustment to the current prediction.
Because its output increment is $p_i h(x_i)$, matching the observations requires $p_i h(x_i)\approx r_i$.
The two update modes therefore motivate different correction
signals for the same residual state.

The additive view uses $r_i$ directly. For the multiplicative view, direct division by $p_i$ would be unstable near zero, so we use the regularized signal:
\begin{equation}
u_i=\frac{p_i r_i}{p_i^2+\lambda^2}.
\label{eq:multiplicative-coordinate}
\end{equation}
Here $\lambda$ is 5\% of the prediction root mean square (RMS), with a positive floor. The dimensionless quantity $u_i$ approximates the fractional adjustment $r_i/p_i$ when $|p_i|$ is large relative to $\lambda$, and suppresses unstable ratios near zero.
Appendix~\ref{app:compact-method-details} gives more details.

The two residual views shown combine the same input coordinates, targets, and predictions with their respective correction signals, $r$ and $u$.
Each view summarizes correlations between transformed coordinates and observation signals. Pairwise statistics capture relationships between variables and the alignment of product corrections with each view's signal. 

The two views share an embedding network and two relation-attention
blocks that encode the variables.
Pair features contribute to both attention biases and values.
For each variable, we fuse its representations from the two views with a pooled global representation, producing $\mathbf h_j$ for variable $j$.
This set-structured design shares parameters across variables and their relationships~\citep{zaheer2017deep,lee2019settransformer}. Padding and masks support inputs with up to eight variables.

\subsubsection{Correction Utility Prediction}

Each correction action selects an additive or multiplicative update, one variable or two distinct variables, and an operator for each selected variable. We use eight operator families:
\begin{equation}
    \Phi=\left\{
        z,\ z^2,\ z^3,\ \sin z,\ \tanh z,\ \exp z,\ \log(1+z^2),\ (1+z^2)^{-1}
    \right\}
\end{equation}
where $z$ denotes a dimensionless scalar argument.
A unary action applies one operator to one variable, while a product action combines two transformed variables.
Masks exclude actions involving inactive variables.

A shared query network combines each variable representation $\mathbf h_j$ with an operator embedding $\mathbf e_k$ to form a variable--operator query.
For product actions, the two queries are combined symmetrically and augmented with their product correlations from both residual views. A shared scoring head combines this representation with the update mode and action type to predict the fitted utility $Q_\theta(\mathcal D,p,a)$.
Appendix~\ref{app:compact-method-details} specifies the action encoding and network architecture.

\subsubsection{Coefficient Fitting and Adaptive Updates}

\paragraph{Coefficient fitting.}
For the selected action, numerical fitting determines the coefficients of its correction function.
For a subset $S$ of observation indices, we define the relative squared-error objective
\begin{equation}
L_S(f)=\frac{1}{|S|}\sum_{i\in S}
\left(\frac{f(x_i)-y_i}{s_{i,S}}\right)^2,
\label{eq:relative-loss}
\end{equation}
where $s_{i,S}$ is the target magnitude with a positive floor scaled to the target RMS on $S$. During refinement, fitting and acceptance use all training observations.

An action with selected variables $J$ and operators $\phi_{k_j}$ defines the correction
\begin{equation}
    h_a(x)=\alpha
    \prod_{j\in J}\phi_{k_j}(b_jz_j+c_j)+\beta,
    \label{eq:correction}
\end{equation}
where $z_j$ is the standardized value of input variable $j$ and $|J|\in\{1,2\}$. The parameters $b_j,c_j$ control each operator's input scale and shift, while $\alpha,\beta$ control the correction's amplitude and offset.
The selected update mode gives either $p+h_a$ or $p(1+h_a)$.
We fit the correction by variable projection: for each setting of the nonlinear parameters $b_j,c_j$, weighted least squares determines the amplitude $\alpha$ and offset $\beta$~\citep{golub1973differentiation,oleary2013variable}. 

\paragraph{Adaptive action selection.}
At each attempt, the frozen scorer selects the highest-scoring untried action. The fitted correction is accepted if it reduces the training relative squared error by more than $0.1\%$. After acceptance, the updated equation determines new residual views and a refreshed action ranking. Rejected actions leave the equation unchanged, and every attempted action is marked as tried. Refinement uses at most four action-fitting attempts, including rejected attempts, and makes no additional LLM calls. Appendix~\ref{app:compact-method-details} gives the point-selection protocol, numerical fitting settings, and early-stopping criterion.

\subsection{Synthetic Data and Model Training}
We train the search residual encoder, structural-edit generator, and refinement model separately using synthetic data generated from elementary functions with sampled transformations, coefficients, and input coordinates, independently of the benchmark equations and observations.
All three components are trained offline, while the pretrained LLM parameters remain frozen. Appendix~\ref{app_training_data} provides more details.

\subsubsection{Search residual Encoder Data and Training}
Target formulas combine randomly transformed elementary functions into compositional products, smooth regime transitions, or condition-gated responses. Current equations omit components or use affine approximations.
Each example pairs a residual state from 256 observations over one to six variables with its target formula.
Validation rejects nonfinite, nearly constant, or excessively large targets and checks numerical alignment and residual consistency.
All target programs pass parsing and compilation.
Audits confirm no cross-split overlap in example identities or exact normalized numerical inputs.
The dataset contains 6,000 training and 768 validation examples.
We train the search residual encoder and projector from scratch by minimizing the LLM loss on target-formula tokens, with the pretrained LLM parameters frozen.

\subsubsection{Structural-Edit Data and Training}
For structural-edit generator training, we generate 20,000 training and 2,000 validation examples, each with 256 observations over one to six variables.
Current formulas omit a target component, contain an extra or incorrect component, or already match the target, providing supervision for addition, deletion, replacement, and stopping.
Numerical evaluation of candidate edits supplies outcome regression and ranking targets.
We train the program encoder and editing modules using supervised edit losses, parameter and outcome regression, and candidate ranking.

\subsubsection{Refinement-Encoder Data and Training}
\paragraph{Refinement-encoder dataset.}
Targets combine three to six unary or product components.
For each target, we construct paired current equations with different residual patterns, covering both coarse and near-target states.
Each pair shares 256 observations and eight candidate actions selected by frozen policies and random sampling.
Sharing observations and actions allows supervision to capture
how an action's utility changes with the current equation.

The splits contain 8,192 training pairs, 1,024 validation pairs, and three test sets of 1,024 pairs covering new observation clouds, held-out product actions, and eight-variable inputs.
Together, they provide 24,576 states and 196,608 action-utility labels.
Validation checks action alignment, variable eligibility, and acceptance decisions.

\paragraph{Fitted-action utility.}
For each synthetic problem, we divide the 256 observations into 192 fitting rows $F$ and 64 utility rows $H$. Rows in $F$ supply the encoder inputs, coefficient fits, and acceptance decisions; rows in $H$ measure the resulting utility. The losses $L_F$ and $L_H$ apply Eq.~\ref{eq:relative-loss} to these respective subsets.

For each current equation and candidate action, we fit the correction
on $F$ and accept it if it reduces $L_F$ by more than $0.1\%$.
Let $p'_a$ denote the accepted equation, or the unchanged parent
when the action is rejected.
Its supervision target is the clipped log reduction in error on $H$:
\begin{equation}
    U(p,a)=\operatorname{clip}_{[-3,3]}
    \left[
    \log_{10}
    \frac{L_H(p)+\varepsilon}{L_H(p'_a)+\varepsilon}
    \right],
    \label{eq:utility}
\end{equation}
where $\varepsilon>0$ stabilizes the logarithm.
Rejected actions have zero utility, while accepted actions receive
positive or negative utility according to their effect on $H$.
The target therefore measures the outcome of fitting a correction
beyond the rows used to estimate its coefficients.

\paragraph{Training objective.}
We jointly train the relational encoder and action scorer. The loss is: 
\begin{equation}
    \mathcal L=\mathcal L_{\mathrm{value}}
    +0.5\mathcal L_{\mathrm{rank}}+0.5\mathcal L_{\Delta}.
    \label{eq:training-loss}
\end{equation}

The value term $\mathcal L_{\mathrm{value}}$ regresses measured utilities. The ranking term $\mathcal L_{\mathrm{rank}}$ learns their ordering within each state~\citep{burges2005learning}. The paired-state term $\mathcal L_{\Delta}$ matches utility differences for the same action across the two current equations. 
We train with AdamW \citep{loshchilov2019adamw} for 16 epochs, corresponding to 2,048 updates, at a learning rate of $3\times10^{-4}$.

\section{Evaluation}
\label{sec:evaluation}

\subsection{Evaluation Setting}
\label{sec:eval_design}

\paragraph{Dataset} 
As many datasets may have been included in LLM pretraining corpora, LLM-SRBench~\citep{shojaee2025bench} combines known terms in the underlying equations with novel, synthetically generated terms to create problems that go beyond memorization. To ensure scientific rigor, two subject-matter experts verified all equations and visualizations of the corresponding generated data.
We evaluate our approach on LSR-Synth, following Xia et al.~\citep{xia2026scientist}. It contains 25 problems in materials science, 36 in chemistry, 24 in biology, and 44 in physics. Each problem contains a training set for SR methods, along with separate in-domain (ID) and out-of-domain (OOD) test sets.


\begin{table}[ht]
\centering
\footnotesize
\setlength{\tabcolsep}{2pt}
\renewcommand{\arraystretch}{0.75} %
\caption{Scientific ID accuracy (\%), measured by $\mathrm{Acc}_{1\%}$ and $\mathrm{Acc}_{0.1\%}$. The best and second-best results are highlighted in bold and underlined, respectively.}
\begin{adjustbox}{width=\linewidth}
\begin{tabular}{@{}l *{10}{c}@{}}
\toprule
\multirow{2}{*}{\textbf{Method}}
    & \multicolumn{2}{c}{\textbf{Overall}}
    & \multicolumn{2}{c}{\textbf{Materials}}
    & \multicolumn{2}{c}{\textbf{Chemistry}}
    & \multicolumn{2}{c}{\textbf{Biology}}
    & \multicolumn{2}{c}{\textbf{Physics}} \\
\cmidrule(lr){2-3}
\cmidrule(lr){4-5}
\cmidrule(lr){6-7}
\cmidrule(lr){8-9}
\cmidrule(lr){10-11}
    & $\mathrm{Acc}_{1\%}$ & $\mathrm{Acc}_{0.1\%}$
    & $\mathrm{Acc}_{1\%}$ & $\mathrm{Acc}_{0.1\%}$
    & $\mathrm{Acc}_{1\%}$ & $\mathrm{Acc}_{0.1\%}$
    & $\mathrm{Acc}_{1\%}$ & $\mathrm{Acc}_{0.1\%}$
    & $\mathrm{Acc}_{1\%}$ & $\mathrm{Acc}_{0.1\%}$ \\
    
\midrule
\multicolumn{11}{c}{\textit{Without LLMs}} \\
\midrule

E2E
    & 0.26 & 0.00
    & 1.33 & 0.00
    & 0.00 & 0.00
    & 0.00 & 0.00
    & 0.00 & 0.00 \\

NeSymReS
    & 3.10 & 0.78
    & 8.00 & 4.00
    & 2.78 & 0.00
    & 4.17 & 0.00
    & 0.00 & 0.00 \\

DSR
    & 0.00 & 0.00
    & 0.00 & 0.00
    & 0.00 & 0.00
    & 0.00 & 0.00
    & 0.00 & 0.00 \\

uDSR
    & 29.46 & 12.40
    & 36.00 & 8.00
    & 50.00 & 25.00
    & 29.17 & 8.33
    & 9.09 & 6.82 \\

PySR
    & 29.46 & 14.47
    & 53.33 & 22.67
    & 25.93 & 11.11
    & 16.67 & 6.95
    & 25.76 & 16.67 \\


\midrule
\multicolumn{11}{c}{\textit{GLM-4.5-Air~\citep{zeng2025glm}}} \\
\midrule

LaSR
    & 14.21 & 8.53
    & 18.67 & 12.00
    & 16.67 & 10.18
    & 11.11 & 5.56
    & 11.36 & 6.82 \\

LLM-SR
    & 35.92 & 14.47
    & 61.33 & 38.67
    & 30.56 & 3.71
    & 18.06 & 4.17
    & 35.61 & 15.15 \\

SR-Scientist
    & 48.32 & 25.06
    & 81.33 & \textbf{70.67}
    & 45.37 & 11.11
    & 40.28 & 16.66
    & 36.37 & 15.15 \\

\midrule
\multicolumn{11}{c}{\textit{Qwen3-Coder-480B-A35B-Instruct~\citep{yang2025qwen3}}} \\
\midrule

LaSR
    & 11.89 & 7.49
    & 13.33 & 6.67
    & 16.67 & 9.26
    & 6.95 & 6.95
    & 9.85 & 6.82 \\

LLM-SR
    & 41.08 & 18.09
    & 80.00 & 52.00
    & 36.11 & 9.26
    & 30.56 & 18.06
    & 28.79 & 6.06 \\

SR-Scientist
    & 49.09 & 24.55
    & \underline{86.67} & \underline{69.33}
    & 40.74 & 5.56
    & \underline{50.00} & \underline{26.39}
    & 34.09 & 13.64 \\

\midrule
\multicolumn{11}{c}{\textit{GPT-OSS-120B~\citep{agarwal2025gpt}}} \\
\midrule

LaSR
    & 16.02 & 10.08
    & 20.00 & 12.00
    & 18.52 & 12.96
    & 9.72 & 5.56
    & 15.15 & 9.09 \\

LLM-SR
    & 28.16 & 11.37
    & 64.00 & 38.67
    & 22.22 & 3.71
    & 11.11 & 2.78
    & 21.97 & 6.82 \\

SR-Scientist
    & \textbf{63.57} & \textbf{49.35}
    & 74.67 & 60.00
    & \textbf{81.48} & \textbf{64.81}
    & \textbf{66.67} & \textbf{43.05}
    & \underline{40.91} & \underline{34.09} \\

\midrule
\multicolumn{11}{c}{\textit{Qwen3-Coder-30B-A3B-Instruct~\citep{yang2025qwen3}}} \\
\midrule

LaSR
    & 12.66 & 8.27
    & 20.00 & 10.67
    & 17.59 & 12.04
    & 8.33 & 6.95
    & 6.82 & 4.55 \\

LLM-SR
    & 24.55 & 7.24
    & 29.33 & 6.67
    & 30.55 & 10.18
    & 15.28 & 5.55
    & 21.97 & 6.06 \\

SR-Scientist
    & 32.30 & 16.02
    & 81.33 & 52.00
    & 22.22 & 5.56
    & 22.22 & 8.33
    & 18.18 & 8.33 \\

SR-Scientist + RL
    & 40.92 & 20.69
    & 85.33 & 65.33
    & 37.38 & 7.46
    & 29.17 & 11.11
    & 25.00 & 11.37 \\

RISR

    & \textbf{63.57} & \underline{38.50}
    & \textbf{89.33} & 58.67
    & \underline{62.04} & \underline{30.56}
    & \underline{50.00} & 25.00
    & \textbf{57.58} & \textbf{40.91} \\
    
\bottomrule
\end{tabular}
\end{adjustbox}
\label{tab:science}
\end{table}

\paragraph{Evaluation Metrics} 
We report accuracy-to-tolerance following the released
SR-Scientist~\citep{xia2026scientist, kamienny2022endtoend, biggio2021neural}.
Point $i$ passes when $|\widehat y_i-y_i|/|y_i| \leq \tau$.
Writing this pointwise indicator as $b_i(\tau)$, task success is
\begin{equation}
\operatorname{Pass}_{\tau}(f)=
\mathbf{1}\!\left[\frac{1}{N}\sum_{i=1}^{N}b_i(\tau) \geq 0.95\right].
\end{equation}
Accuracy is the percentage of tasks that pass, reported at $\tau=0.01$ and $0.001$.

\begin{figure}[ht]
    \centering
    \includegraphics[width=\linewidth]{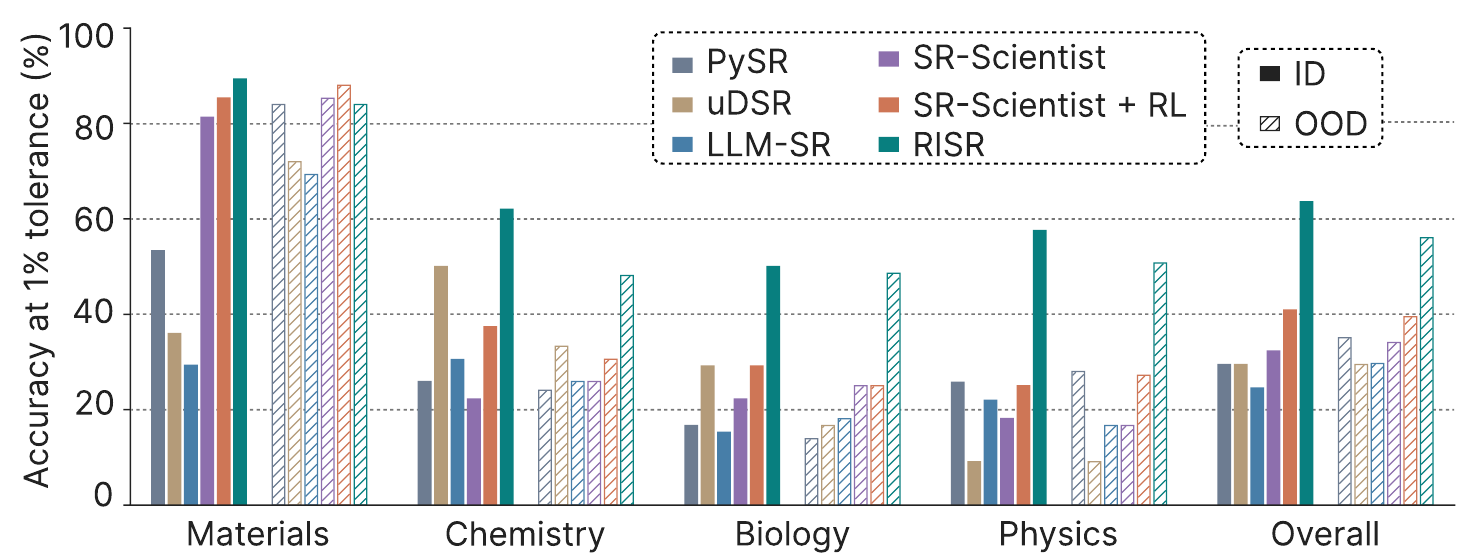}
    \caption{ID and OOD accuracy at a $1\%$ relative-error tolerance. Colors denote methods. Filled and hatched bars represent ID and OOD results, respectively. All LLM-based methods use Qwen3-Coder-30B-A3B-Instruct.
    }
    \label{fig:ood}
\end{figure}

\paragraph{Baselines}
We compare RISR against both non-LLM and LLM-based symbolic regression methods. The non-LLM baselines include E2E~\citep{kamienny2022endtoend}, NeSymReS~\citep{biggio2021neural}, DSR~\citep{brenden2021deep}, uDSR~\citep{landajuela2022unified}, and PySR~\citep{cranmer2023pysr}, while the LLM-based baselines comprise LaSR~\citep{grayeli2024lasr}, LLM-SR~\citep{shojaee2025llmsr}, and SR-Scientist~\citep{xia2026scientist}. We use the published results reported by~\citep{xia2026scientist}. 

\paragraph{Implementation and Protocol}
We use Qwen3-Coder-30B-A3B-Instruct as the search backbone. Its pretrained parameters remain frozen during encoder training and evaluation. 
Each search has 64 outer candidate slots, including initialization. LLM proposals and learned structural edits share this budget.
Stage B first fits the selected equation's coefficients and then allows at most four new correction-fitting attempts, including rejections, without further LLM calls.
Within-run equation selection and fitting use training observations, leaving the test set for evaluation.

\subsection{Main Results}
\label{sec:scientific_results}
\paragraph{ID Performance} 
Table~\ref{tab:science} reports the ID performance of \name{}, averaged across results of three seeds. More details can be found in Appendix~\ref{app:result_statistics}.
Among methods using the same backbone, \name{} achieves 63.57\% and 38.50\% accuracy at the 1\% and 0.1\% tolerances, respectively, exceeding SR-Scientist+RL by 22.65 and 17.81 percentage points. The gains span all four domains at the 1\% tolerance, with particularly large improvements in chemistry and biology: 24.66 and 20.83 percentage points, respectively.
This breadth indicates that the overall improvement is shared across scientific domains rather than concentrated in a single problem category.

The 30B model results are also competitive with systems built on substantially larger backbones. 
\name{} exceeds the strongest reported Qwen3-Coder-480B result in seven of the ten dimensions. 
At the 1\% tolerance, across all baselines, it matches the highest reported overall
accuracy, $63.57\%$ for SR-Scientist with GPT-OSS-120B.
At the stricter 0.1\% tolerance, its 38.50\% accuracy ranks second overall, ahead of all reported GLM-4.5-Air and Qwen3-Coder-480B configurations.
These results demonstrate the effectiveness of the complete \name{} pipeline at the 30B scale and show that leading accuracy at the 1\% tolerance can be achieved without adopting a larger LLM backbone.

As for domains, the strongest cross-backbone results occur in materials science and physics.
\name{} achieves the highest materials accuracy at the 1\% tolerance, reaching 89.33\%.
In physics, it leads at both tolerances with 57.58\% and 40.91\%, exceeding the strongest competing results of 120B model by 16.67 and 6.82 percentage points, respectively. 
In chemistry, RISR ranks second at both tolerances. In biology, it ties
for second place at the 1\% tolerance.

\paragraph{OOD performance}
\name{} achieves the highest overall OOD accuracy among the compared methods at the same 30B model scale (Figure~\ref{fig:ood}).
At the 1\% tolerance, its accuracy reaches 56.07\%, compared with 39.50\% for SR-Scientist + RL, a gain of 16.57 percentage points. 
At $1\%$, OOD accuracy is $84.00\%$ in materials science, $48.15\%$ in chemistry, $48.61\%$ in biology, and $50.76\%$ in physics. Physics retains $40.91\%$ accuracy at the stricter $0.1\%$ tolerance.
These results show that the advantage of \name{} extends beyond ID fitting to prediction outside the training domain using the same discovered equations.

\subsection{Ablation Study}
\label{sec:scientific_stages}

\begin{table}[t]
\centering
\small
\setlength{\tabcolsep}{4pt}
\caption{The ablation results with one seed setting. Accuracy is reported at the $1\%$ / $0.1\%$ tolerances. For Stage B, all rows share the same Stage-A initial equations and numerical fitting protocol.}
\label{tab:science_stages}
\begin{tabularx}{\textwidth}{@{}Xrr@{}}
\toprule
Configuration & ID accuracy $\uparrow$ & OOD accuracy $\uparrow$ \\
\midrule
LLM-only formula search & 7.75 / 0.00 & 9.30 / 3.10 \\
+ residual-informed formula search (Stage A)  & 36.43 / 23.26
& 40.31 / 30.23 \\
\quad+ coefficient fitting & 59.69 / 34.11
& 53.49 / 38.76 \\
\quad\quad+ random corrections & 60.47 / 34.88
& 52.71 / 36.43 \\

\quad\quad+ learned-scored corrections (RISR) & 63.57 / 37.21 & 
55.04 / 38.76 \\
\bottomrule
\end{tabularx}
\end{table}

Table~\ref{tab:science_stages} isolates the contributions of coefficient fitting and the learned correction-selection policy.
The complete residual-informed search raises accuracy at the 1\% tolerance from 7.75\% to 36.43\% on ID data and from 9.30\% to 40.31\% on OOD data compared with the LLM-only method.
Residual-guided refinement provides additional gains, reaching 63.57\% ID and 55.04\% OOD accuracy at the 1\% tolerance.
Compared with random corrections, learned-scored corrections improve ID and OOD accuracy at the 1\% tolerance by 3.10 and 2.33 percentage points.
The results show a positive contribution from residual-informed formula search and learned action selection under the fixed correction budget.

\subsection{Symbolic Accuracy}
\label{sec:symbolic_accuracy}

\begin{wraptable}{r}{0.4\textwidth}
\centering\small
\setlength{\tabcolsep}{5pt}
\caption{SA (\%) on one seed setting.}
\label{tab:symbolic_accuracy}
\begin{tabularx}{\linewidth}{@{}l>{\raggedright\arraybackslash}Xr@{}}
\toprule
Method & LLM backbone & SA $\uparrow$ \\
\midrule
uDSR & --- & 0.77 \\
PySR & --- & 4.65 \\
LLM-SR & GLM-4.5-Air & 5.43 \\
LLM-SR & GPT-OSS-120B & 4.65 \\
LLM-SR & Qwen3-Coder-480B & 5.43 \\
SR-Scientist & GLM-4.5-Air & 7.75 \\
SR-Scientist & GPT-OSS-120B & 7.00 \\
SR-Scientist & Qwen3-Coder-480B & 7.00 \\
\midrule
RISR & Qwen3-Coder-30B & \textbf{13.18} \\
\bottomrule
\end{tabularx}

\end{wraptable}

Symbolic accuracy (SA) evaluates whether a discovered equation can represent the target relationship. We follow the symbolic-assessment criterion of SR-Scientist~\citep{xia2026scientist}: A candidate is counted as positive when an assignment of its adjustable constants can represent the target expression. GPT-6-Astra serves as our judge, and non-unanimous cases are reserved for human review. 17 problems receive ten Yes votes, and 112 receive ten No votes, leaving no voting disagreements. We additionally manually reviewed all 17 cases classified as positive by the LLM judge.
RISR obtains an SA of $13.18\%$ ($17/129$), exceeding existing methods (Table~\ref{tab:symbolic_accuracy}).
SA and numerical accuracy are complementary: predictive agreement on held-out observations does not necessarily imply structural recovery under the constant-assignment criterion.
Appendix~\ref{app:symbolic_accuracy_protocol} documents more details.

\subsection{Case Study}
\label{sec:case-po30}

\begin{figure}[t]
\centering
\includegraphics[width=\linewidth]{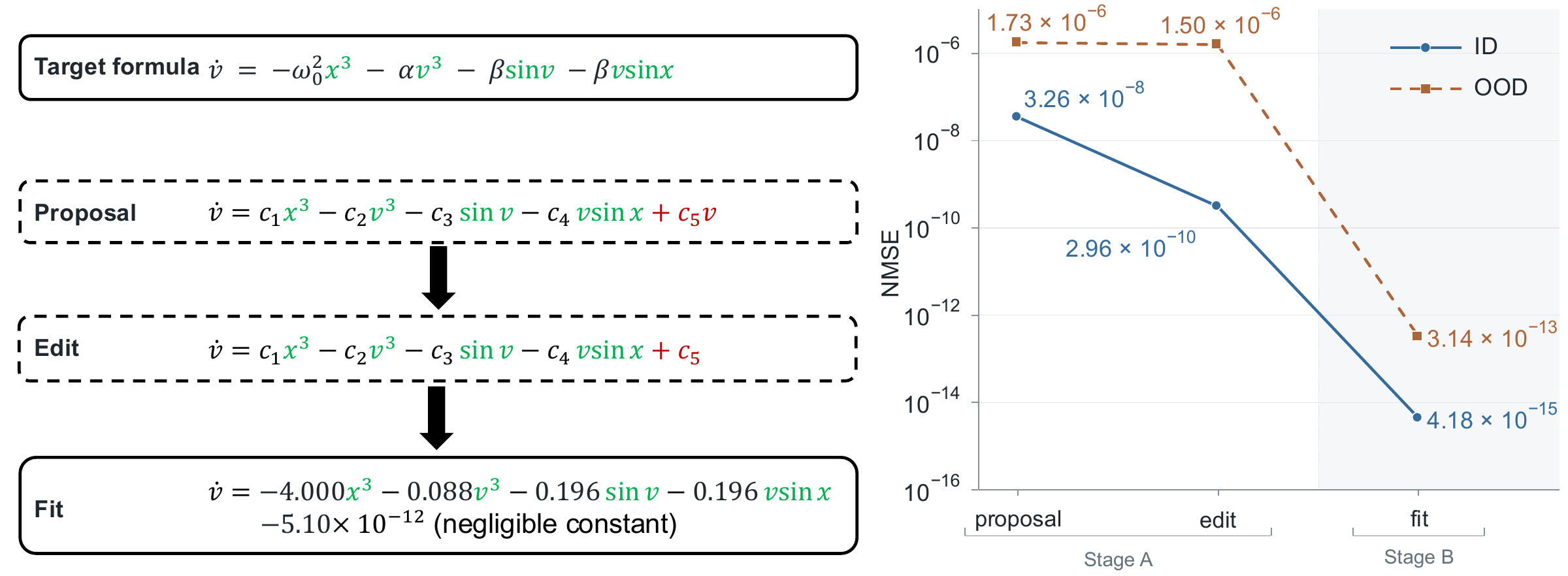}
\caption{Case study on PO30. \name{} retains the target components, removes an extraneous term through structural editing, and calibrates the coefficients during refinement.}
\label{fig:case-po30}
\vspace{-12pt}
\end{figure}

PO30 illustrates how RISR contributes to equation recovery (Figure~\ref{fig:case-po30}).
The target describes a nonlinear oscillator, with position $x$ and velocity $v$.
During Stage A, the formula proposal contains all four target components, but also includes an extra term $c_5 * v$.
Then the learned structural-edit generator replaces this extra term with a constant and submits the edited formula for coefficient fitting and scoring.
At Stage B, coefficient fitting calibrates the selected equation without changing its structure.
The remaining constant becomes negligible, and the two independently fitted sinusoidal coefficients closely match the shared coefficient
in the target.

We examine more cases (Appendix~\ref{app:qualitative-trajectories}), including those in which numerical error is substantially reduced despite incomplete recovery of the underlying formula. These cases may inform future research aimed not only at reducing numerical error but also at recovering the scientific formulas.

\section{Conclusion}

We present \name{}, a two-stage method that uses the pointwise residual information of a current equation to guide scientific equation discovery.
Residual-conditioned LLM proposals and structural-edit generators support formula search, while a dual-view encoder predicts which corrections are likely to improve the equation after fitting.
For the evaluation, \name{} reaches 63.57\% and 38.50\% ID accuracy at the 1\% and 0.1\% tolerances.
The case study illustrates how structural editing and coefficient calibration contribute at different points in the discovery process.
The evaluation results show the effectiveness of our method in the task of scientific equation discovery.

\subsection*{AI Use Statement}
In this work, we used generative AI tools to assist in the writing of proofs, implement methods, assist with translation, and formulate mathematical claims.
We have not used generative AI tools to generate synthetic data sets, propose or refine hypotheses, design or provide feedback on research methodology or experiments, clean and reformat datasets, support qualitative and thematic data analysis, or interpret results.
Providing critical ingredients for proving mathematical claims and helping develop theoretical models or conceptual frameworks is not applicable to this work.
We have reviewed all AI-assisted work. We take responsibility for the final content of this work, including text, claims, or artifacts produced with the aid of generative AI.

\section*{Ethics Statement}
\name{} is developed and evaluated using publicly available scientific benchmark data and synthetic numerical datasets generated through transformations and compositions of elementary mathematical functions. This work focuses on methodological research in scientific equation discovery. We have not identified specific ethical concerns associated with the data and experiments presented in this study.
\section*{Reproducibility Statement}
The main paper and appendices describe the model architectures, synthetic data generation procedures, training settings, and evaluation protocols used in our experiments. To facilitate reproducibility, we will publicly release the synthetic datasets and the code for data generation, model training, equation search and refinement, and evaluation.

\bibliography{iclr2027_conference}
\bibliographystyle{iclr2027_conference}

\appendix
\section{Details of Residual-Informed Formula Search}
\label{app:upstream-encoder}
This section describes the configuration used for \name{}. Stage A combines residual-conditioned LLM proposals and learned structural edits. 

\subsection{Formula search and numerical inputs}
The search extends the evaluator--buffer--sampler design of LLM-SR~\citep{shojaee2025llmsr}. Executable candidates are fitted to the training observations and stored with their training MSE, predictions, and residuals. The buffer maintains ten islands and samples two equation programs from one island to form each prompt.
The highest-scoring equation included in that prompt supplies the parent predictions for numerical conditioning. The textual input contains an equation-completion instruction, variable descriptions, and the sampled code context. Pointwise observations are supplied through continuous embeddings rather than a natural-language account of the residuals.

The residual encoder receives aligned coordinates $x_i$, targets $y_i$, parent predictions $p_i$, and residuals $r_i=y_i-p_i$. It has six coordinate slots. For a problem with more than six variables, each coordinate is scored by its largest absolute correlation with the residual, considering raw, absolute-value, and squared transformations on both sides. The six highest-scoring coordinates are retained in their original order, with coordinate-index tie breaking. Routing affects only this encoder: the equation signature and numerical fitting retain all original variables. The search configuration allows up to 4,000 encoding rows. If subsampling is required, rows are sorted by residual and one aligned row is sampled from each equal-count stratum.

Normalization statistics are computed on the retained rows. Coordinate $j$ is centered by its median and divided by the larger of its interquartile range and standard deviation. A scale at most $10^{-8}$ is replaced by one. Coordinates are clipped to $[-12,12]$ and padded to six slots. Let $\mu_y$ be the target mean and $\sigma_y=\max\{\operatorname{std}(y),10^{-8}\}$. Targets and predictions are centered by $\mu_y$ and divided by $\sigma_y$.
Residuals are divided by $\sigma_y$ without centering. These three channels are clipped to $[-20,20]$. The normalized residual's magnitude and signed $\log(1+|r|)$ complete five observation channels. Each row therefore has eleven entries, with a separate mask for active coordinate slots.

\subsection{Point-attention and spectral-field paths}
\begin{figure}[tbp]
\centering
\includegraphics[width=\linewidth]
{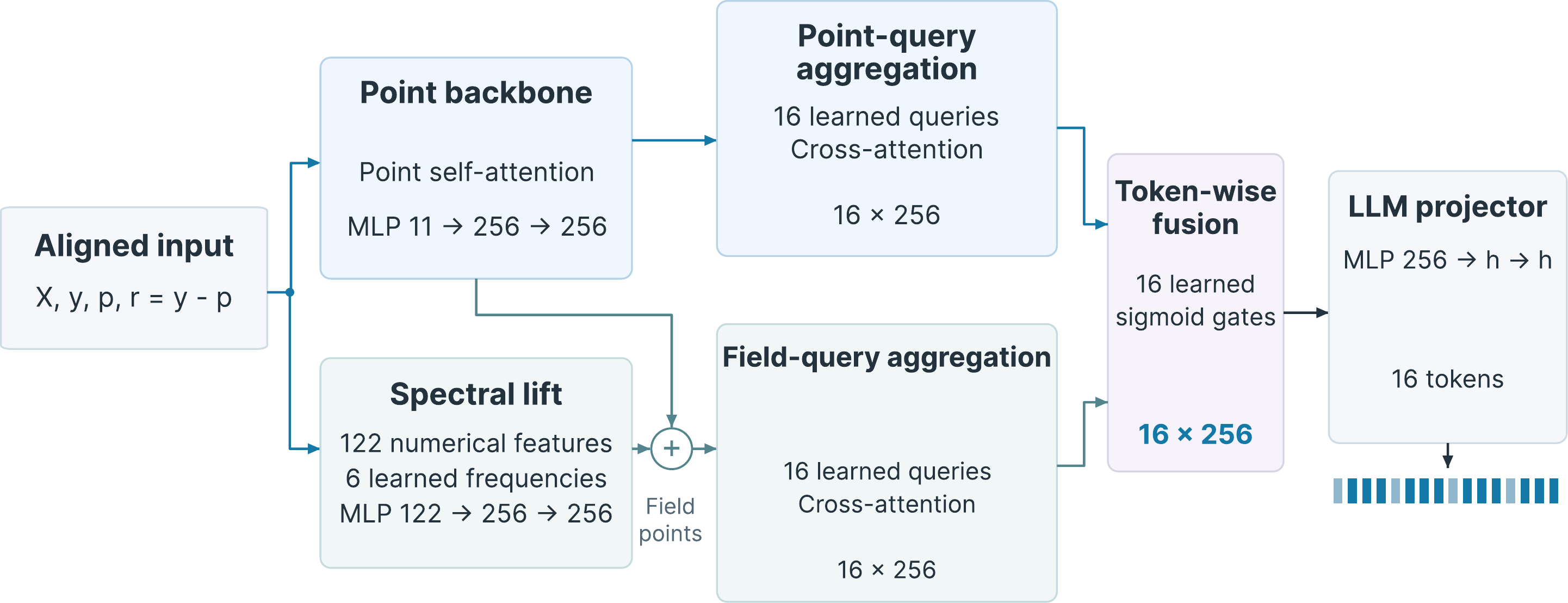}
\caption{Residual encoder architecture.}
\label{fig:app-search-encoder}
\end{figure}

Figure~\ref{fig:app-search-encoder} shows how the residual encoder maps aligned numerical observations to continuous LLM tokens through point-attention and spectral-field paths.
\paragraph{Point-attention path.}
A two-layer $11\!\to\!256\!\to\!256$ MLP embeds each row.
Two eight-head self-attention blocks mix point features, followed by
layer normalization. For large observation sets, attention operates
within chunks after sorting by normalized residual plus a small
deterministic projection of the full row. Search-time chunks contain
at most 256 points. Sixteen learned queries aggregate the valid points
through two cross-attention blocks. Their shared context combines an
MLP of feature means, standard deviations, and the variable mask with
an MLP of the mask alone. A final layer normalization yields sixteen
vectors of width 256.

\paragraph{Spectral-field path.}
The second path adds an explicit feature lift to the point
representations. It contains 24 coordinate channels (each coordinate
alone and multiplied by the normalized residual, its magnitude, or
its square), 21 coordinate-pair products multiplied by the residual,
72 sine/cosine channels, and the five observation channels. Thus,
each point supplies $24+21+72+5=122$ features. The coordinate-pair
channels include squares. Six positive frequencies, shared across
coordinates, are learned in log space and initialized on a
logarithmic grid from one to eight. Their sinusoidal arguments
include a factor of $\pi$. Features associated with padded
coordinates are masked. An MLP maps the lift to width 256 and adds
it to the point features. These are learned frequency features,
not an FFT of the observations.

Seven learned residual centers, initialized uniformly from $-3$ to
$3$, define additional summaries. For each center, Gaussian-shaped
weights depend on normalized residual distance and a shared learned
positive precision. Weights sum to one over valid points. A weighted
field-feature average and a learned region embedding form each
summary. Sixteen field queries attend to both point features and
the seven summaries. Two cross-attention/latent-self-attention pairs,
followed by layer normalization, produce sixteen width-256 vectors.

\paragraph{Fusion and projection.}
One learned sigmoid gate per token interpolates between the two
paths. The gates are shared across examples. A two-layer GELU
projector with widths $256\!\to\!2048\!\to\!2048$, layer
normalization, and a learned positive output scale maps the fused
vectors to the Qwen embedding space. The scale starts at $0.02$.
One active marker embedding is replaced by the sixteen continuous
vectors, and the sequence mask is updated. Inserted positions do
not receive target-formula token labels. The LLM is frozen during
encoder training, and both networks operate in evaluation mode
during equation search.

\subsection{Auxiliary learned structural-edit generator}
\label{app:structural-editor}
\begin{figure}[tbp]
\centering
\includegraphics[
width=0.9\linewidth,
height=0.78\textheight,
keepaspectratio]
{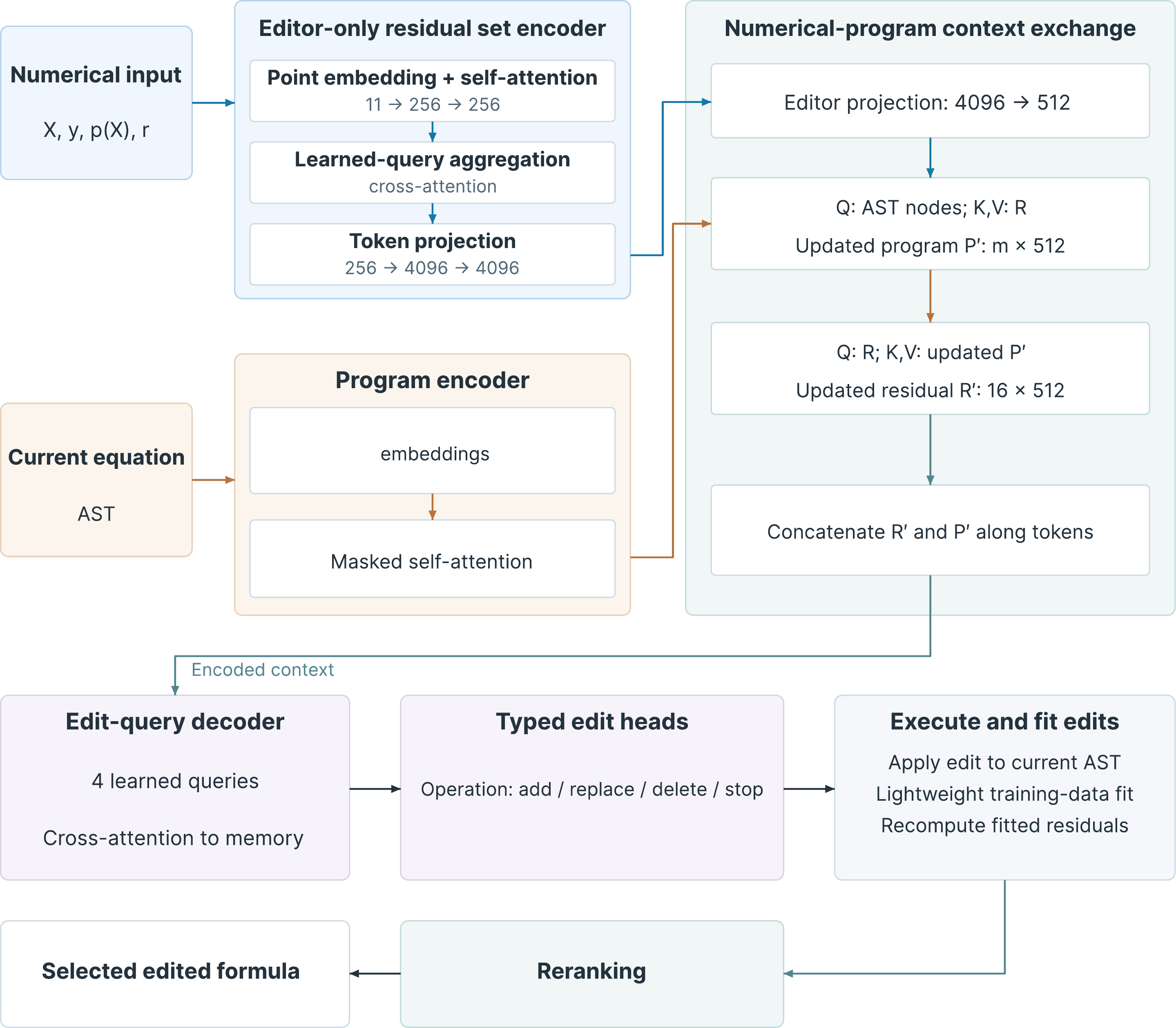}
\caption{Numerical and program encoders within the structural edit generator.}
\label{fig:app-editor-encoders}
\end{figure}

Figure~\ref{fig:app-editor-encoders} shows the structural editor's numerical and program encoders.
The editor takes the training observations and the current formula.
Its own residual encoder summarizes coordinates, targets,
predictions, and residuals into sixteen latent vectors, projected to
the editor's hidden width. A program encoder represents the formula's
abstract syntax tree, including node types, variables, constants,
parent relations, and depths. Cross-attention combines the numerical
and program representations to predict an edit operation, target node,
function family, variable support, and continuous parameters. The
available operations are addition, replacement, deletion, and stopping.
The input program is limited to 96 nodes and numerical encoding uses
at most 512 observations.

Predicted edits are converted into executable formulas and evaluated
with lightweight training-data fitting. Their resulting residuals
inform learned outcome scores used for reranking. A selected edit
is then submitted to the common outer evaluator, which fits its
coefficients and scores its training MSE. The editor's predicted
extrapolation-related score is learned from synthetic data. It does
not evaluate the benchmark OOD split during search. All editor
weights are frozen at deployment. Its pretraining and edit supervision
are described in Appendix~\ref{app:editor-data}.

\subsection{Stage-A LLM Calling}

LLM decoding uses temperature $0.8$, top-$k=30$, top-$p=0.9$,
repetition penalty $1.0$, and at most 256 new tokens per completion.
The buffer cluster temperature is $0.1$. Outer coefficient fitting
uses BFGS with one configured restart and a 120-second evaluation
timeout. Among executable logged candidates that reproduce their
recorded training fit, the equation with the lowest training MSE is
selected, retaining the earlier candidate on ties. This selected
equation is the Stage-A output $p_0$.

\section{Details of Residual-Guided Refinement}
\label{app:compact-method-details}

\subsection{Dual-view construction and relational statistics}
\begin{figure}[tbp]
\centering
\includegraphics[width=\linewidth]
{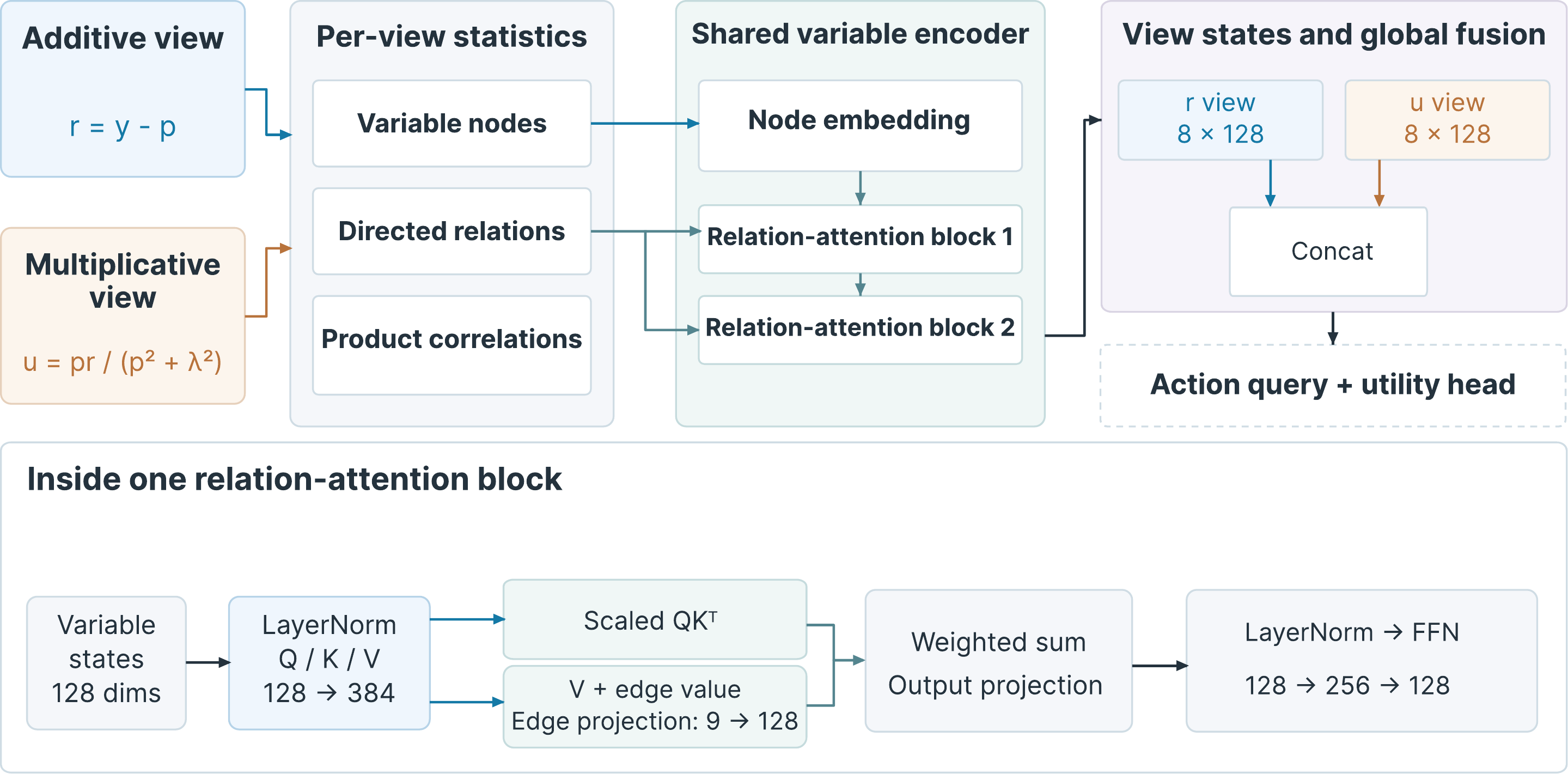}
\caption{Dual-view refinement encoder.
The two views share the embedding network and relation-attention
weights.
The lower panel shows how pair features modify attention
logits and values.}
\label{fig:app-refinement-encoder}
\end{figure}

Figure~\ref{fig:app-refinement-encoder} details the dual-view refinement encoder and its relation-attention blocks. The encoder maps additive and multiplicative residual statistics to variable representations used by the correction utility scorer.

For predictions $p_i$ and residuals $r_i=y_i-p_i$, the regularized
multiplicative coordinate solves
\begin{equation}
u_i=\arg\min_{u\in\mathbb R}\{(r_i-p_i u)^2+\lambda^2u^2\}
    =\frac{p_i r_i}{p_i^2+\lambda^2},\qquad
\lambda=0.05\max\{\operatorname{RMS}(p),10^{-8}\}.
\label{eq:app-multiplicative}
\end{equation}
Prediction RMS is computed on the encoding rows. The positive floor
makes $u_i$ defined even when all predictions vanish. The additive
view uses $r_i/\sigma_y$; the multiplicative view uses
$u_i/\max\{\operatorname{std}(u),10^{-8}\}$, without centering
the signal. Both signals are clipped to $[-20,20]$.

Coordinates use median/IQR-or-standard-deviation normalization and
clipping as in the residual encoder, but are padded to eight slots.
Each view also contains target and prediction channels normalized
by the target mean and standard deviation, its correction signal,
the signal magnitude, and its signed logarithm. Squaring and cubing
the signal yields seven observation features. For a normalized
coordinate $z$, the fourteen coordinate transforms are
\begin{equation}
\begin{split}
\Psi(z)=\big[&z,z^2,z^3,|z|,\log(1+z^2),e^{-z^2},\\
&\{\sin(\pi f z),\cos(\pi f z)\}_{f\in\{0.5,1,2,4\}}\big].
\end{split}
\end{equation}
Cross-correlations of the fourteen transforms with seven signals,
plus their means and log standard deviations, provide
$14\times7+2\times14+2\times7=140$ features per variable.
Moment standardization floors the variance at $10^{-8}$.

For each directed variable pair $(j,k)$, correlations of
$z_jz_k$, $z_j^2z_k$, and $z_jz_k^2$ with the correction signal,
its magnitude, and its square give nine edge features. Separately,
each unordered pair has 64 operator-product correlations in each
view. These use the eight correction operators evaluated on
normalized coordinates, standardize each operator response, and
correlate their products with the view signal. Operator-response
and signal standard deviations are floored at $10^{-4}$; product
variances are floored at $10^{-8}$, and correlations are clipped
to $[-1,1]$. Exponentials in this statistical probe use arguments
clipped to $[-5,5]$. This probe clipping does not redefine the
exponential operator in fitted correction formulas.

A shared $140\!\to\!128\!\to\!128$ embedding and two four-head
relation-attention blocks encode each view. Linear maps of the
nine edge features contribute attention biases and values.
Each block has residual connections and a width-$256$ feed-forward
hidden layer. Masked mean pooling of the two views supplies a
global width-$128$ context. Concatenating this context with each
variable's two representations gives a width-$384$ vector, which
a $384\!\to\!128$ fusion layer maps to $\mathbf h_j$.

\subsection{Action vocabulary and shared utility head}

Each action specifies an additive or multiplicative update, one
variable or two distinct variables, and one operator per variable.
The operator order is
\begin{equation}
\Phi=\{z,z^2,z^3,\sin z,\tanh z,\exp z,
         \log(1+z^2),(1+z^2)^{-1}\}.
\end{equation}
There are $16d+128\binom d2$ valid actions at dimension $d$, or
3,712 for eight variables. Product factors are stored in canonical
variable order, and actions involving padded variables are masked.
A shared network maps the concatenation of $\mathbf h_j$ and a
16-dimensional operator embedding to a 64-dimensional query
$\mathbf q_{j,k}$. The action representation is
\begin{equation}
\begin{split}
\mathbf v_a=\big[&\mathbf q_L+\mathbf q_R,\;
\mathbf q_L\odot\mathbf q_R,\;|\mathbf q_L-\mathbf q_R|,\\
&\mathbf e_m,\;I_{\mathrm{product}},\;
c_a^{(+)},\;c_a^{(\times)}\big],
\end{split}
\end{equation}
where $\mathbf e_m$ is a 16-dimensional mode embedding and the
last two scalars are the action's product correlations. Unary
actions set $\mathbf q_R=0$, $I_{\mathrm{product}}=0$, and both
correlations to zero. A shared $211\!\to\!128\!\to\!1$ head
predicts fitted utility. The refinement encoder, queries, and
utility head together contain 389,297 trainable parameters.

\subsection{Correction coefficient fitting}
For an observation subset $S$, the fitting loss is
\begin{equation}
L_S(f)=\frac1{|S|}\sum_{i\in S}
\left(\frac{f(x_i)-y_i}{s_{i,S}}\right)^2,\qquad
s_{i,S}=\max\{|y_i|,0.01\operatorname{RMS}(y_S),\epsilon_{64}\},
\label{eq:app-relative-loss}
\end{equation}
where $\epsilon_{64}$ is double-precision machine epsilon. This
weighted squared-error objective is different from the test-time
accuracy-to-tolerance criterion.

At the start of Stage B, all adjustable coefficients of $p_0$ are
fitted on the full training set with its structure fixed. The
trust-region reflective least-squares solver starts from the saved
coefficients, uses a two-point numerical Jacobian, at most 500
function evaluations, and $10^{-10}$ for each termination tolerance.
Parameter scales are $\max\{|c_j|,1\}$. A finite strict improvement
in $L_S$ is retained; otherwise the original coefficients are kept.
This initial calibration is separate from the four-attempt correction
budget and does not use the $0.1\%$ minimum-improvement threshold
required for accepting a new correction.

An action defines $h_a=\alpha g_a+\beta$, where
$g_a(x)=\prod_{j\in J}\phi_{k_j}(b_j z_j+c_j)$ and
$z_j=(x_j-\mu_j)/\max\{\sigma_j,10^{-8}\}$ uses fitting-row
means and standard deviations. For fixed nonlinear parameters,
the two linear design columns are $(g_a,1)$ for an additive update
and $(p g_a,p)$ for a multiplicative update, with response $y-p$.
Dividing the design and response by $s_{i,S}$ gives the weighted
least-squares problem for $\alpha,\beta$. The current equation's
existing coefficients remain fixed during every new correction fit.

The nonlinear parameters use three deterministic starts. Unary
starts for $(b,c)$ are $(1,0)$, $(-1,0)$, and $(2,0.5)$; product
starts for $(b_1,c_1,b_2,c_2)$ are $(1,0,1,0)$,
$(-1,0,1,0)$, and $(2,0.5,0.5,-0.5)$. Slopes are bounded by
$[-20,20]$ and shifts by $[-10,10]$. Each start uses trust-region
reflective least squares, a two-point Jacobian, and tolerances of $10^{-9}$. Each action has a
15-second overall deadline across its starts. The best finite fit
completed within the budget is retained. An attempt with no finite
candidate leaves the equation unchanged; solver convergence alone
is not the acceptance criterion.

\subsection{Encoding rows, action selection, and stopping}

The current refinement configuration encodes 256 training rows.
When more rows are available, a deterministic sample without
replacement is drawn using seed 20260831 and its indices are
sorted. The same indices are used after accepted updates; their
predictions, residuals, and normalization statistics are recomputed.
This fixed encoding-row seed is distinct from the Stage-A search
seed. Every coefficient fit and acceptance decision uses the full
training set. During synthetic utility training, encoding and fitting
instead use the designated 192 fitting rows.

At each attempt, the frozen model ranks dimension-valid actions and
selects the highest-scoring untried action, using a stable order for
ties. For the fitted equation $\widetilde p_a$, acceptance requires
finite training predictions and
$L_S(\widetilde p_a)<0.999L_S(p)$. An accepted action updates $p$
and its residual state; a rejection retains $p$. Every attempted
action is marked as tried, and all attempts count toward the cap of
four. Refinement stops before a new attempt when training NMSE is
at most $10^{-12}$. Thus, a task may require zero new correction
fits even though its initial coefficients were refitted. No Stage-B
operation calls the LLM or uses test observations for acceptance.

\section{Synthetic Data and Offline Training}
\label{app_training_data}
The two main training corpora supervise residual-to-formula generation
and fitted correction utility. The auxiliary structural editor has
additional residual-pretraining and edit-supervision datasets.
All are generated from elementary functions, sampled transformations,
coefficients, and coordinates. They do not use benchmark equations
or observation arrays as generation inputs. Table~\ref{tab:app-data}
separates their sizes and learning roles.

\begin{table}[htbp]
\centering
\caption{Synthetic training collections. Counts are examples except
for the paired utility corpus, whose counts are pairs. The three
utility test sets are additional to its training and validation splits.}
\label{tab:app-data}
\begin{tabularx}{\textwidth}{@{}XrrX@{}}
\toprule
Collection & Train & Validation & Supervision \\
\midrule
Search residual-to-formula & 6,000 & 768 & Target formula tokens \\
Editor residual pretraining & 50,000 & 5,000 & Families, variables, gains \\
Structural edits & 20,000 & 2,000 & Edits and fitted outcomes \\
Paired correction utility & 8,192 & 1,024 & Eight actions per state \\
\bottomrule
\end{tabularx}

\end{table}

\subsection{Search encoder data and training}

Three families contribute 2,000 training and 256 validation examples
each. Compositional examples multiply two to six transformed factors
with a global amplitude and at most ten target-program parameters;
the current equation omits a factor or uses an affine approximation.
Smooth-regime examples combine coordinate and condition responses
through sigmoid or $\tanh$ gates; the current equation omits the
late regime or uses an affine approximation. Factorized responses
combine onset, hardening or softening, condition-dependent factors,
and a smooth damage gate; current equations omit a condition factor
or damage gate, or use an affine approximation.

Each example has 256 observations and one to six variables.
Compositional coordinates are sampled uniformly from randomly
chosen intervals among $[-3,3]$, $[0.05,5]$, $[0,1]$, and $[0,40]$.
The other families use two randomly assigned active coordinates
in $[0,1]$, with additional coordinates in $[-1,1]$. Elementary
transformations include shifts, rescaling, powers, saturation,
exponential decay or onset, squared sine, and hinges. Generation
rejects nonfinite targets, target standard deviation at most
$10^{-4}$, and absolute outputs exceeding $10^5$ for compositional
examples or $10^4$ for the other families. Validation checks row
alignment, finite inputs, and $r=y-p$. All 6,768 target programs
pass parsing and compilation; this is a syntactic validation and
is not a proof of identifiability from observations.

The residual encoder and projector are randomly initialized and
trained through target-formula token cross-entropy, with all LLM
weights frozen and no LoRA adapter. Training uses AdamW, one epoch
over 6,000 examples, an effective batch size of eight, and 750
updates. Learning rates are $2\times10^{-5}$ for the encoder trunk
and $10^{-4}$ for the projector, with weight decay $0.01$. 

\subsection{Structural-editor data and training}
\label{app:editor-data}
The editor's residual encoder is separately pretrained on 50,000
training and 5,000 validation examples. A sampled sum of elementary
terms forms a current equation; additional sampled terms form the
clean target. The library includes constants, powers, trigonometric
and saturating functions, bounded exponentials, protected inverses
and ratios, and variable products. Labels identify function
families, active variables, and numerical fitting gains. This
encoder is frozen during subsequent edit training.

Edit training uses 20,000 training and 2,000 validation examples,
each with 256 observations and one to six variables. Current
programs omit a target component, contain an extra or incorrect
component, or already match the target. These constructions provide
addition, deletion, replacement, and stop labels. Up to four valid
candidate slots provide measured fitting gains, gains on separately
sampled wider-range synthetic inputs, and changes in expression
complexity. Gaussian observation noise has standard deviation
equal to $0.5\%$ of the clean-target standard deviation, with a
positive floor. Structural-record hashes separate training and
validation.

The program encoder and editing modules learn supervised operation,
location, family, and variable predictions, continuous parameters,
outcome regressions, and candidate rankings. AdamW training uses
ten epochs, batch size sixteen, learning rate $2\times10^{-4}$,
and weight decay $0.01$. Selection uses validation loss. The
synthetic wider-range observations provide offline supervision;
they are not the benchmark OOD arrays.

\subsection{Paired-state refinement data}

Targets contain three to six unary or two-variable product
components using the eight correction operators. Affine input
slopes are sampled in $[0.6,2.2]$ and shifts in $[-1,1]$.
Components are combined by addition, multiplication by one plus
a component, saturation through $\tanh$, or division by one plus
a component's square. An affine output transformation varies scale
from $10^{-3}$ to $10^3$. Each problem has 256 points drawn from
uniform, clipped-normal, or standardized-lognormal distributions,
followed by coordinate rescaling and translation. Training and
validation use two to six variables; the dimensional test uses
eight. 

Two current equations share each observation cloud and noisy target.
Coarse states are formed by removing an additive component,
dividing out a multiplicative component, fitting a subset of
components, or combining these operations. One quarter of pairs
remain coarse. For the others, each prediction is rescaled toward
the clean target,
\begin{equation}
p^{(s)}=f_{\mathrm{clean}}+\alpha_s
        (p^{(s)}_{\mathrm{coarse}}-f_{\mathrm{clean}}),
\qquad s\in\{0,1\},
\end{equation}
placing its clean-error contribution, measured on the first 192
rows, into one of eight logarithmic bins spanning $10^{-8}$ to
$1$. Observation noise is unchanged. The encoder receives the
numerical state, not the target expression.

Each pair shares eight dimension-valid actions. Frozen earlier
residual-based policies propose top actions for both current
states, and a frozen target-only policy supplies additional
candidates. After deduplication, random valid actions fill the
set to eight, retaining at least two random candidates.

\begin{table}[htbp]
\centering
\caption{Paired-state utility splits. Each pair has two states and
eight actions per state. Only the training split contributes
labels to gradient updates.}
\label{tab:utility_corpus_splits}
\begin{tabularx}{\textwidth}{@{}Xrrr@{}}
\toprule
Split & Pairs & States & Utility labels \\
\midrule
Training & 8,192 & 16,384 & 131,072 \\
Validation & 1,024 & 2,048 & 16,384 \\
New observation clouds & 1,024 & 2,048 & 16,384 \\
Held-out product actions & 1,024 & 2,048 & 16,384 \\
Eight-variable inputs & 1,024 & 2,048 & 16,384 \\
\midrule
Total & 12,288 & 24,576 & 196,608 \\
\bottomrule
\end{tabularx}

\end{table}

Splits use separate random streams and keep both states of a pair
together. In the held-out-product test, operator-index pairs with
$(o_1+o_2)\bmod4=0$ are reserved for candidate actions and excluded
from candidates in the other splits. Target-expression construction
can still use all operator combinations. This tests held-out
correction actions rather than disjoint target-function families.
Validation checks shared-action alignment, variable eligibility,
acceptance decisions, and utility arithmetic. Positive, zero,
and negative labels are retained.

\subsection{Utility targets, losses, and optimization}

The first 192 rows form $F$ and supply encoder inputs, coefficient
fits, and acceptance decisions. The remaining 64 rows form $H$
and measure fitted utility. Some target-generation normalizations
use the full synthetic cloud before this assignment; $H$ is
therefore held out from correction fitting, not from every
operation used to construct the synthetic problem. Neither subset
contains benchmark test observations. After fitting on $F$, define
\begin{equation}
p'_a=\begin{cases}
\widetilde p_a,&L_F(\widetilde p_a)<0.999L_F(p)
                 \text{ and the fit is finite},\\
p,&\text{otherwise}.
\end{cases}
\end{equation}
The supervision target is
\begin{equation}
U(p,a)=\operatorname{clip}_{[-3,3]}
\left[\log_{10}\frac{L_H(p)+10^{-8}}{L_H(p'_a)+10^{-8}}\right].
\label{eq:app-utility}
\end{equation}
The same loss in Eq.~\ref{eq:app-relative-loss} is evaluated
separately on $F$ and $H$. Rejection gives zero utility; nonfinite
utility-row predictions receive $-3$. An accepted action may have
negative utility if it worsens prediction on $H$.

Let $\widehat U_{s,a}$ be the predicted utility and $\ell$ the
smooth-$L_1$ loss with transition at absolute error one. The value
term averages $\ell(\widehat U_{s,a}-U_{s,a})$. For actions $a,b$
in the same state, define $d_{s,ab}=U_{s,a}-U_{s,b}$,
$\widehat d_{s,ab}=\widehat U_{s,a}-\widehat U_{s,b}$, and
$w_{s,ab}=\min\{|d_{s,ab}|,3\}\mathbf1[|d_{s,ab}|>0.05]$.
The ranking loss is
\begin{equation}
\mathcal L_{\mathrm{rank}}=
\frac{\sum_{s,a,b}w_{s,ab}
 \log(1+\exp[-\operatorname{sign}(d_{s,ab})\widehat d_{s,ab}])}
 {\max\{1,\sum_{s,a,b}w_{s,ab}\}}.
\end{equation}
For each paired current state and shared action, the paired term
averages
\begin{equation}
\ell\big((\widehat U_{0,a}-\widehat U_{1,a})
            -(U_{0,a}-U_{1,a})\big).
\end{equation}
The total is $\mathcal L_{\mathrm{value}}+
0.5\mathcal L_{\mathrm{rank}}+0.5\mathcal L_\Delta$.

The refinement encoder, action-query network, and utility head are
trained jointly from random initialization. AdamW uses learning
rate $3\times10^{-4}$, weight decay $0.01$, and gradient-norm
clipping at one. A batch contains 64 pairs, or 128 states. Sixteen
epochs over 8,192 pairs give 2,048 updates. The selected model remains frozen during refinement.

\section{Result statistics}
\label{app:result_statistics}
The main table and ID/OOD plot average three complete results with seeds 0, 42, and 20260831 (Table~\ref{tab:snapshot_search_runs}).
We first compute
the success rate for each seed, then take the mean of
the three rates. Standard deviations are sample standard deviations
with two degrees of freedom. Domain results use the corresponding
domain's task count.

\begin{table}[htbp]
\centering
\caption{Detailed results for the three seeds.}
\setlength{\tabcolsep}{3pt}
\begin{tabularx}{\textwidth}{@{}Xrrrr@{}}
\toprule
Result set & ID, 1\% & ID, 0.1\% & OOD, 1\% & OOD, 0.1\% \\
\midrule
Seed 0 & 65.89 (85) & 42.64 (55) & 57.36 (74) & 41.09 (53) \\
Seed 42 & 61.24 (79) & 35.66 (46) & 55.81 (72) & 34.88 (45) \\
20260831 & 63.57 (82) & 37.21 (48) & 55.04 (71) & 38.76 (50) \\
\midrule
Mean $\pm$ SD & $63.57\pm2.33$ & $38.50\pm3.66$ & $56.07\pm1.18$ & $38.24\pm3.13$ \\
\bottomrule
\end{tabularx}

\label{tab:snapshot_search_runs}
\end{table}

\section{Symbolic Accuracy: Protocol and Detailed Results}
\label{app:symbolic_accuracy_protocol}
SA uses the constant-assignment criterion of
SR-Scientist~\citep{xia2026scientist}: whether some assignment
of the candidate's adjustable constants makes it represent the
target relationship. The original program parameters and the
parameters of accepted corrections remain adjustable for this
assessment. Coordinate normalization, operator choices, powers,
clipping guards, and fixed literals remain unchanged. 

SA uses all 129 final programs and includes every accepted correction. With their saved coefficients, the exported programs reproduce stored predictions in all 387 training/ID/OOD replay checks. 

GPT-6-Astra evaluates each candidate--target pair in ten separate
requests. The recorded requests specify high reasoning effort and
a limit of 8,192 completion tokens, without an explicit temperature
or sampling seed. The prompt asks about the existence of suitable
constant assignments and requests a structured Yes/No answer.
Unanimous Yes and No outcomes are accepted; non-unanimous
outcomes are reserved for human review. The completed assessment
has 1,290 valid judgments: 17 tasks receive ten Yes votes and
112 receive ten No votes, with no unresolved cases (Table~\ref{tab:sa_domains}).

\begin{table}[htbp]
\centering
\caption{Symbolic accuracy for the seed 20260831. All positive
tasks receive ten Yes votes; all remaining tasks receive ten No votes.}
\begin{tabularx}{\textwidth}{@{}Xrrr@{}}
\toprule
Domain & Positive & Tasks & SA (\%) \\
\midrule
Materials & 1 & 25 & 4.00 \\
Chemistry & 5 & 36 & 13.89 \\
Biology & 3 & 24 & 12.50 \\
Physics & 8 & 44 & 18.18 \\
\midrule
Overall & 17 & 129 & 13.18 \\
\bottomrule
\end{tabularx}

\label{tab:sa_domains}
\end{table}

\section{More Case Studies}
\label{app:qualitative-trajectories}

The examples below report saved formulas, selected actions,
acceptance decisions, and replayed errors. The residual encoder supplies
continuous numerical tokens, and the Stage-B model outputs utility
scores rather than textual reasoning.

\subsection{CRK22: a four-step numerical refinement}
\label{app:traj-crk22}

The chemistry reference has the structure
\begin{equation}
\dot A=-kA^2-kA\exp(-k_s t)
       +\frac{k_z A^2}{1+qA^4},
\end{equation}
where $t$ is time and $A$ is concentration. 

\begin{table}[htbp]
\centering
\caption{CRK22 action sequence. All four actions are additive and
accepted by the training objective. Define
$\operatorname{rq}(z)=(1+z^2)^{-1}$ and
$\operatorname{lsq}(z)=\log(1+z^2)$. }
\label{tab:app-crk22}
\begin{tabularx}{\textwidth}{@{}rlXr@{}}
\toprule
Attempt & Mode & Correction family & OOD NMSE \\
\midrule
0 & --- & Fitted parent & $2.0917$ \\
1 & Additive & $\operatorname{rq}(t)\,\tanh(A)$ & $4.0358\times10^{-4}$ \\
2 & Additive & $\operatorname{rq}(t)\,\operatorname{lsq}(A)$ & $2.9398\times10^{-4}$ \\
3 & Additive & $\operatorname{rq}(t)\,\sin(A)$ & $1.7143\times10^{-4}$ \\
4 & Additive & $\sin(t)\,\operatorname{rq}(A)$ & $6.4102\times10^{-6}$ \\
\bottomrule
\end{tabularx}

\end{table}

ID NMSE changes from $1.6757\times10^{-7}$ to
$4.7072\times10^{-8}$. The OOD point fraction below each of
the two main tolerances changes from zero to one. Each accepted
correction changes the residual state used for the next ranking.
The final formula is nevertheless not counted as symbolically
equivalent by the ten-vote assessment.

\subsection{MatSci9: additive and multiplicative corrections}
\label{app:traj-matsci9}

MatSci9 (Table~\ref{tab:app-matsci9}) alternates between multiplicative and additive actions
on strain $\epsilon$ and temperature $T$. A multiplicative
action means $p\leftarrow p(1+h_a)$, not multiplication by the
displayed operator product alone. Every listed factor again
has fitted affine parameters, and all four attempts are accepted.

\begin{table}[htbp]
\centering
\caption{MatSci9 refinement trajectory. Action names omit fitted
affine parameters, amplitudes, and offsets.}
\label{tab:app-matsci9}
\begin{tabularx}{\textwidth}{@{}rlXr@{}}
\toprule
Attempt & Mode & Correction family & OOD NMSE \\
\midrule
0 & --- & Fitted parent & $5.478\times10^{-7}$ \\
1 & Multiplicative & $\operatorname{cube}(\epsilon)\operatorname{square}(T)$ & $1.471\times10^{-8}$ \\
2 & Multiplicative & $\tanh(\epsilon)\operatorname{linear}(T)$ & $3.322\times10^{-9}$ \\
3 & Additive & $\operatorname{lsq}(\epsilon)\operatorname{lsq}(T)$ & $2.878\times10^{-9}$ \\
4 & Multiplicative & $\operatorname{cube}(T)$ & $7.566\times10^{-9}$ \\
\bottomrule
\end{tabularx}

\end{table}

ID NMSE decreases from $1.490\times10^{-9}$ to
$2.580\times10^{-12}$. Final OOD NMSE is lower than the
parent's, but higher than at the third prefix. The final OOD
point fraction is one at both main tolerances. This nonmonotonic
test error is consistent with training-only acceptance and is
retained rather than replaced by the best test-scoring prefix.

\section{Utility-Ranking Experiments}
\label{app:utility-ranking-diagnostic}
Each synthetic example contains two current equations sharing the same
observation cloud and the same eight candidate actions. Let $s$ denote
one state and $\bar{s}$ its paired state. For the paired-state swap
diagnostic, we use the scorer outputs produced for $\bar{s}$ to select
an action for $s$:
\[
a_s^{\mathrm{swap}}
=
\arg\max_{a}
\widehat U_{\bar{s},a},
\qquad
U_s(a_s^{\mathrm{swap}})
\text{ is then evaluated using the utility labels of }s.
\]
The same operation is performed in the opposite direction for
$\bar{s}$. Thus, the action pool and fitting protocol are unchanged;
only the predicted scores are taken from the other current equation.

\begin{table}[t]
\centering\small
\caption{Held-out utility of the selected action in fixed eight-action candidate pools. Each column contains 2,048 states. Utilities are clipped base-10 log error reductions measured after coefficient fitting. Higher values are better.}
\label{tab:review_pool_utility}
\begin{tabularx}{\linewidth}{@{}lrrr@{}}
\toprule
Selector & New observation clouds & Unseen product actions
& Eight-variable inputs \\
\midrule
Learned utility scorer & 0.46273 & 0.40994 & 0.45324 \\
Residual-correlation ranking & 0.39336 & 0.37877 & 0.39710 \\
Uniform random selection & 0.24101 & 0.22907 & 0.19557 \\
Paired-state swap & 0.27099 & 0.25677 & 0.22226 \\
\bottomrule
\end{tabularx}

\end{table}
To directly test whether the learned scorer identifies promising corrections, we evaluate it on three held-out synthetic splits with a fixed pool of eight dimension-valid actions per state. 
Each split contains 1,024 paired problems, or
2,048 current-equation states. Every state is associated with a fixed
pool of eight dimension-valid correction actions. 

The refinement model is frozen during this evaluation. For every
state and candidate action, the correction is fitted on the designated
fitting rows using the same variable-projection procedure as in the
refinement stage. The fitted equation is accepted only when it reduces
the fitting loss by more than $0.1\%$. Its utility is then measured on
the disjoint utility rows using the clipped logarithmic error reduction
defined in Eq.~\ref{eq:utility}. The reported score is the mean utility
of the action selected by each selector.

The results are shown in Table~\ref{tab:review_pool_utility}. The learned scorer achieves mean utilities of $0.4627$, $0.4099$, and $0.4532$ on new observation clouds, unseen product actions, and eight-variable inputs, respectively. The learned scorer outperforms residual-correlation ranking and
uniform random selection on all three held-out splits. The
paired-state swap is substantially worse, which is consistent with
the scorer using information about the current equation rather than
only the shared action pool. This result provides direct evidence that the learned utility scorer can identify more promising corrections than simple selection rules in controlled candidate pools.

\section{Computational Accounting}
\label{app:computational-accounting}

\paragraph{Search budget.}
Stage A permits at most 64 outer candidate submissions, including the
initial seed equation. Invalid submissions also consume an outer slot.
After the outer search terminates, executable candidates that reproduce
their recorded training fit are compared using the training objective,
and the candidate with the lowest training MSE is selected as $p_0$.
Stage B first fits the coefficients of $p_0$ under the relative-error
objective. This calibration is separate from the subsequent budget of
at most four new correction-fitting attempts.

\paragraph{Measured cost.}
The Stage-A configuration generates four continuations per LLM
request. A complete 64-slot task search therefore uses at most 64 LLM generation requests.

For the seed 20260831, in Stage B, the refinement logs contain 424 new
correction attempts, of which 400 are accepted. The projected fitting
objective is evaluated 461,860 times, and the recorded correction-fit
durations sum to 433.98 seconds.

\end{document}